%% file: main.tex
\documentclass[a4paper, conference]{IEEEtran}
\IEEEoverridecommandlockouts
\usepackage{cite}
\usepackage{amsmath,amssymb,amsfonts}
\usepackage{algorithmic}
\usepackage{graphicx}
\usepackage{textcomp}
\usepackage{siunitx}
\usepackage{tikz}
\usetikzlibrary{arrows}
\usepackage{amsmath}

\usepackage{xcolor}
\definecolor{todo}{rgb}{1,0.5,0}

\usepackage{pgfplots}
\usepackage{caption}
\usepackage{subcaption}

\usepackage[a4paper, left=1.29cm,right=1.29cm,top=1.8cm,bottom=4.30cm]{geometry}

\def\BibTeX{{\rm B\kern-.05em{\sc i\kern-.025em b}\kern-.08em
    T\kern-.1667em\lower.7ex\hbox{E}\kern-.125emX}}
    
\newcommand\copyrighttext{%
  \footnotesize \textcopyright 2023 IEEE. Personal use of this material is permitted.
  Permission from IEEE must be obtained for all other uses, in any current or future
  media, including reprinting/republishing this material for advertising or promotional
  purposes, creating new collective works, for resale or redistribution to servers or
  lists, or reuse of any copyrighted component of this work in other works.
  }
\newcommand\copyrightnotice{%
\begin{tikzpicture}[remember picture,overlay]
\node[anchor=south,yshift=10pt] at (current page.south) {\fbox{\parbox{\dimexpr\textwidth-\fboxsep-\fboxrule\relax}{\copyrighttext}}};
\end{tikzpicture}%
}   
    
\begin{document}

\title{LEO-PNT With Starlink: Development of a Burst Detection Algorithm Based on Signal Measurements
\thanks{The authors are with the SpaceCom Labs of the Space Systems Research Center, Bundeswehr University Munich, 85579 Neubiberg, Germany (e-mail:  papers.sp@unibw.de). This research is partly funded by dtec.bw – Digitalization and
Technology Research Center of the Bundeswehr. dtec.bw is funded by the European Union - NextGenerationEU.}
}
\author{\IEEEauthorblockN{Winfried Stock, Christian A. Hofmann, and Andreas Knopp}
\IEEEauthorblockA{\textit{Institute of Information Technology, University of the Bundeswehr Munich, Neubiberg, Germany} \\
papers.sp@unibw.de}
}

\maketitle

\copyrightnotice

\begin{abstract}
Due to the strong dependency of our societies on Global Navigation Satellite Systems and their vulnerability to outages, there is an urgent need for additional navigation systems. A possible approach for such an additional system uses the communication signals of the emerging LEO satellite mega-constellations as signals of opportunity. The Doppler shift of those signals is leveraged to calculate positioning, navigation and timing information. Therefore the signals have to be detected and the frequency has to be estimated. In this paper, we present the results of Starlink signal measurements. The results are used to develope a novel correlation-based detection algorithm for Starlink burst signals. The carrier frequency of the detected bursts is measured and the attainable positioning accuracy is estimated. It is shown, that the presented algorithms are applicable for a navigation solution in an operationally relevant setup using an omnidirectional antenna.
%
\end{abstract}

\begin{IEEEkeywords}
LEO, Starlink, navigation, signals of opportunity 
\end{IEEEkeywords}

\section{Introduction}
At least since the widespread use of smartphones, Positioning, Navigation and Timing (PNT) functionalities have become a matter of course in the everyday lives of very many citizens. Additionally, in the context of Industry 4.0 the importance of PNT has been increasing drastically for industry, agriculture, etc. in the last few years. Most devices that offer PNT functionality rely on Global Navigation Satellite Systems (GNSS). Due to some deficiencies of GNSS, e.g., when used in cities or forests and in jamming scenarios, there is an urgent need for additional systems. Plans for including PNT functionalities in the future 6G standard for cellular networks underline this need.

A promising approach for an additional navigation system uses Signals of Opportunity (SoO)\nocite{b4}. Such systems use signals that are not designed or transmitted for the purpose of navigation but make secondary use of already available signals. The advantage of an SoO approach is that no dedicated transmitters have to be operated. On the downside, the structure of the utilized signals is not optimized for PNT and is often unknown.

In the previous decades, especially terrestrial signal sources, such as cellular basestations, were considered as SoO. With the New Space movement and the exponentially growing number of LEO satellites in operation, the signals of LEO satellites have become more and more interesting as a source for an SoO navigation approach (LEO-PNT). Several characteristics of LEO satellites make them a promising source: There are LEO satellite signals available all the time on the whole globe, and the orbits of the satellites are known through the TLE files published by the North American Aerospace Defense Command (NORAD). Additionally, LEO satellites move with high velocity (relative to, e.g., a terrestrial receiver), which (over time) causes a fast changing geometry. Furthermore, a considerable Doppler shift is caused, which can be leveraged for PNT. In such an approach, PNT information is derived from frequency measurements of the received signal (similar to the measurement of the time of arrival of GNSS signals and the calculation of PNT information by triangulation).





This paper focuses on using Starlink signals as SoO for a LEO-PNT system. 
Since 2018, the Starlink system has been continously growing and is now the constellation with the largest number of operational satellites in the LEO \cite{space2022}. Hence, Starlink is a suitable candidate to provide SoO for LEO-PNT.

However, until recently, very little was known about the structure of Starlink signals. For this reason, so far, the published opportunistic LEO-PNT implementations have leveraged rather superficial signal properties, like the bandwidth, peaks in the frequency domain, and an assumed periodicity of the signal \cite{Neinavaie2021, Neinavaie2022b, Neinavaie2022a, Khalife2022}.
Due to the recent publication of an in-depth analysis of the Starlink user downlink signal in \cite{Humphreys2022}, dedicated Starlink signal detection and frequency estimation algorithms can be developed that utilize the exact signal structure and might offer a significantly higher estimation accuracy.






In this work, the basic working principles of Doppler shift based LEO-PNT are described. The Starlink user uplink signal is measured and analyzed. To the knowledge of the authors, this work is the first one doing so. Similarities between the uplink and downlink signal structures are identified.
Algorithms for burst detection and frequency estimation that utilize the Starlink synchronization sequence are proposed and analyzed.
Finally, the impact of the frequency estimation errors of those algorithms on the positioning accuracy of Doppler shift based LEO-PNT is investigated by calculating its lower bound.


\section{Opportunistic LEO-PNT} \label{secLEO-PNT}

\begin{figure}[!t]
\centering
\setlength{\abovecaptionskip}{-10pt}    
\input{figBlockDiagramReceiver.tex}
\caption{Generic receiver architecture for Doppler shift based LEO-PNT}
\label{figBlockDiagramReceiver}
\end{figure}
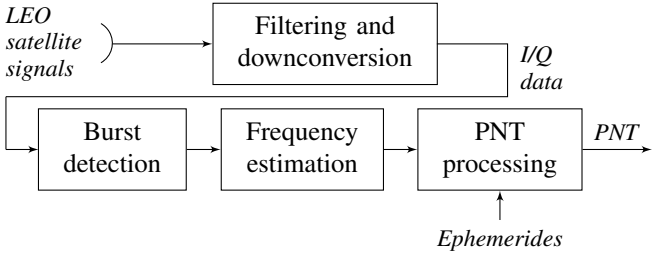


The basic working principle of an opportunistic LEO-PNT receiver leveraging the Doppler shift of LEO communication
signals is depicted in figure \ref{figBlockDiagramReceiver}. After an initial filtering and downconversion, the LEO burst signal is detected, and a carrier frequency estimation is conducted. Subsequently, PNT information is calculated from the received carrier frequency $f_r$ and the location and velocity of the satellite. This calculation is possible because the high relative velocity between the transmitting satellite and the receiver causes a significant Doppler shift $f_D$, which depends on the relative position between the satellite and receiver. The velocity and location of the satellite are usually obtained from the ephemerides made public by NORAD in TLE files \cite{NORAD}. Strategies to identify the individual satellite (to match the transmitting satellite and the TLE file) have to be applied. As the signal properties of individual satellites are usually not known publicly, those strategies are, for example, based on preknowledge of the approximate receiver position. In the following, this section provides a better understanding of PNT calculation from frequency measurements.
\nocite{Middlestead2017}

\begin{figure}[!t]
\centering
\includegraphics[scale = 0.37]{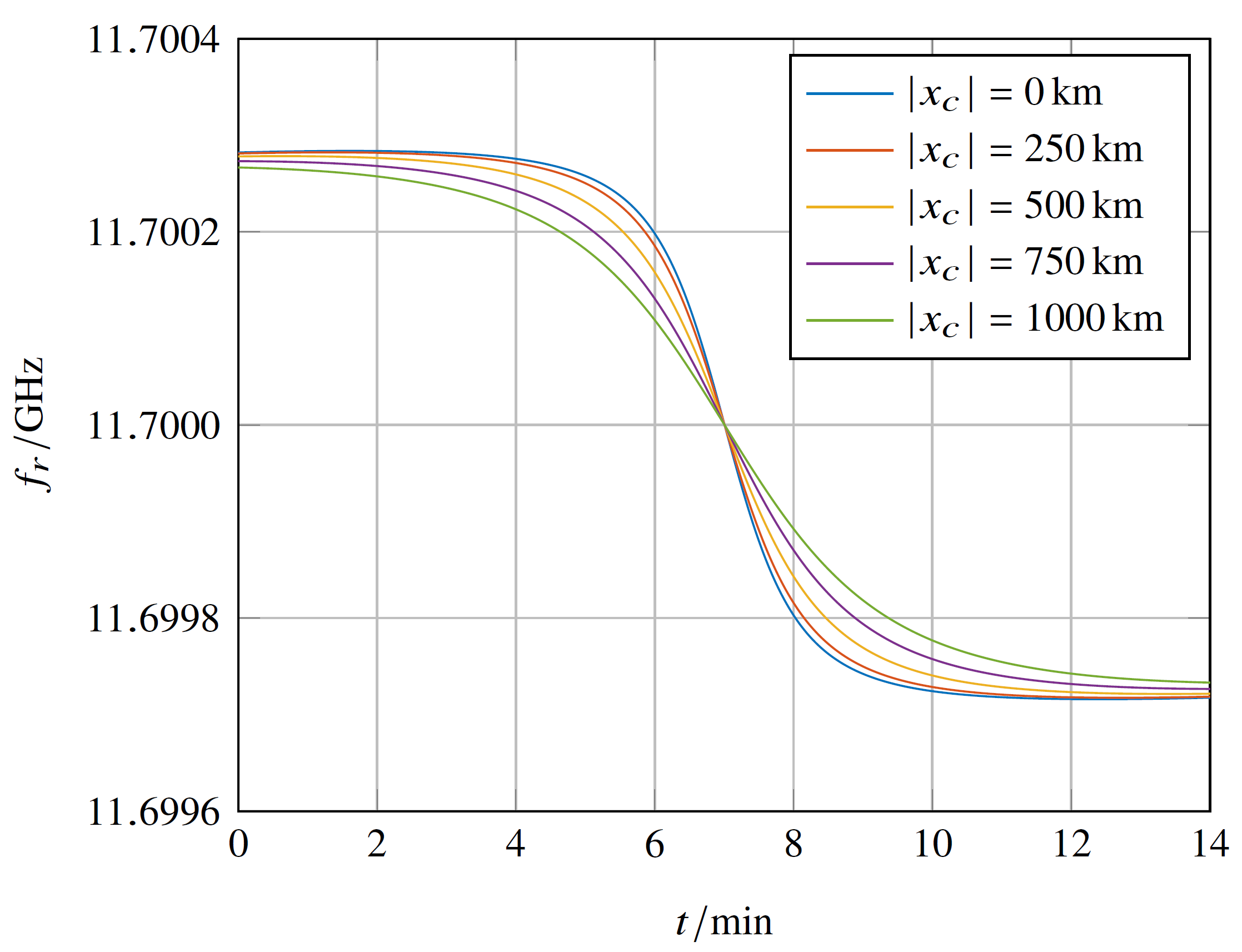}
\caption{Received center frequency $f_r$ for different receiver positions in cross-track direction during a satellite overflight}
\label{figDopplerCurve}
\end{figure}

During the overflight of a satellite, the changing relative velocity between transmitter and receiver causes a characteristic evolution of the Doppler shift over time. The precise shape of this Doppler shift curve depends on the receiver position relative to the satellite trajectory.  Figure \ref{figDopplerCurve} shows some examples for different locations in cross-track direction $\vec{x}_c$ of a static receiver on the earth surface. Different locations in along-track direction $\vec{x}_a$ produce the same Doppler shift curves, but time-shifted. ($\vec{x}_a$ refers to the direction parallel to the ground track of the satellite; $\vec{x}_c$ is perpendicular to $\vec{x}_a$ and parallel to the earth surface.) Mathematical models to calculate the Doppler shift can be found, e.g., in \cite{Psiaki2021}.

When the signal of a single satellite is tracked over the timespan $t_a$, commonly, an opportunistic LEO-PNT receiver estimates (at least some of) the following parameters: the receiver's 3D-position, the receiver's 3D-velocity, the receiver time, and the transmitted carrier frequency. The last parameter can also include the (possibly time-varying) clock drifts of the satellite and the receiver. The number of conducted frequency measurements $N$ must at least match the number of parameters that are estimated by the receiver. However, the higher the number of measurements, the more accurate the PNT estimation is. The same is true for $t_a$. A longer tracking duration $t_a$ entails measurements with a greater variety of geometry between satellite and receiver, which improves accuracy. For the same reason, the PNT estimation accuracy can be significantly improved by conducting measurements from several satellites with different orbits.
 
Several sources introduce errors to Doppler shift based PNT estimation. Among the most significant sources are the following three. First, the orbit of the satellite is not known exactly. The TLE files published by NORAD entail satellite position errors of up to a few kilometers \cite{Kassas2021}. Second, errors are introduced by the clocks of the receiver and of the satellite. The latter cannot be assumed to have an accuracy comparable to those of GNSS. (This, e.g., results in time varying carrier frequency offsets or sampling frequency offsets.) Last, the frequency estimation introduces errors. While the first-mentioned sources are not in the scope of this work, Section \ref{positioningAccuracy} will focus on the impact of the last-mentioned error source on the positioning accuracy.

\section{Starlink signal measurements} \label{secSignal}
This section provides an overview of the Starlink signals. At first, a signal model is presented that fits the user uplink as well as the user downlink signal. After that, the uplink signal is described based on signal measurements and analysis conducted in the course of this work. Finally, the downlink signal is described shortly based on the in-depth analysis provided very recently by \cite{Humphreys2022}. 

\subsection{Signal model}
Due to similarities in their structure, the same signal model can be applied for Starlink uplink and downlink signals:
\begin{equation}
s_i[n] = \alpha (c_i[n] + d_i[n-L_c]) e^{j2\pi (f_{ci} + f_{Di}) T_s n} + w[n]
\end{equation}  
where $s_i[n]$ is the $i$-th received signal burst at the $n$-th time instant. The data signal in baseband is denoted by $d_i[n]$, the noise term by $w[n]$, and $T_s = \frac{1}{f_s}$ denotes the sampling period. The carrier frequency $f_{ci}$ depends on the used subchannel and, therefore, can vary from burst to burst. 
The received signal bursts are shifted in frequency by $f_{Di}$, which includes the Doppler shift as well as (at least for uplink signals) the Doppler shift pre-compensation applied by the transmitter. Within each burst, $f_{Di}$ is assumed to be constant. The complex channel gain $\alpha$ describes all channel effects except the Doppler shift. The baseband synchronization sequence $c_i[n]$ with length $L_c$ and $n \in \{0, ..., L_c-1 \}$ can be described by
\begin{equation}
c_i[n] =  \dot{c}^p + \sum_{k=0}^{7} \dot{c}^k_i \big[n-(k+\gamma)\dot{L}_c\big]
\end{equation}
where $\dot{c}^k_i$ denotes the $k$-th subsequence with $k \in \{0,1,...,7\}$. Each subsequence has a length of $\dot{L}_c$ samples. The relationship between those elements can be described with $-\dot{c}_{i}^0 = \dot{c}_{i}^1 = \dot{c}_{i}^2 = ... = \dot{c}_{i}^7$. The prefix $\dot{c}_{i}^p$ is a cyclic prefix of $\dot{c}_{i}^0$ and consists of $\gamma \dot{L}_c$ samples. As $c_i$ is assumed to be the same for each burst, the index $i$ can be omitted.

\subsection{Uplink signal properties}
\paragraph*{Conducted measurements}

\begin{figure}[!t]
\centering
\setlength{\abovecaptionskip}{10pt}    
\input{figBlockDiagramMeasurement.tex}
\caption{Signal measurement setup}
\label{figBlockDiagramMeasurement}
\end{figure}
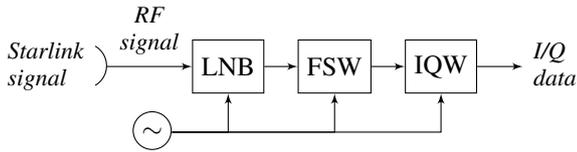

\begin{figure}[!t]
\includegraphics[scale=0.45, clip]{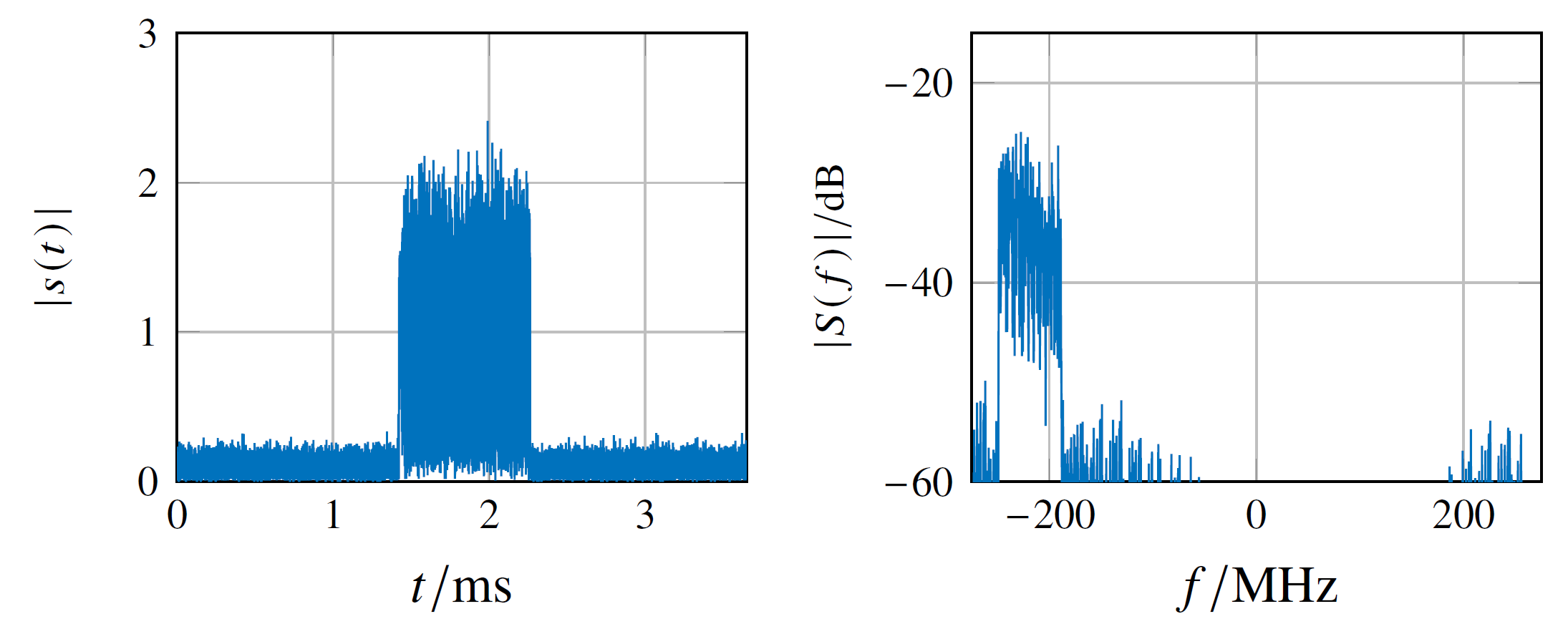}
\caption{Exemplary Starlink uplink burst in time and frequency domain in baseband}
\label{figBurstOne}
\end{figure}

In the course of this work, Starlink user uplink signal measurements were conducted with the setup depicted in figure \ref{figBlockDiagramMeasurement}. A horn antenna was placed next to an active Starlink user terminal. The received signal was amplified and downconverted in a low noise block downconverter (LNB). A R\&S\textsuperscript{\textregistered}FSW spectrum analyzer and R\&S\textsuperscript{\textregistered}IQW wideband I/Q data recorder were used to downconvert the signal to (quasi) baseband and store it. The equipment was synchronized by a \SI{10}{\mega\hertz} rubidium oscillator. The stored signal was analyzed in MATLAB\textsuperscript{\textregistered}. 
The investigated signal has a duration of \SI{80}{\second} and a sampling rate of $f_s = 562.500.000\ \si{\second^{-1}}$. 

\begin{figure}[!t]
\includegraphics[trim={0 0 3mm 0},scale=0.44]{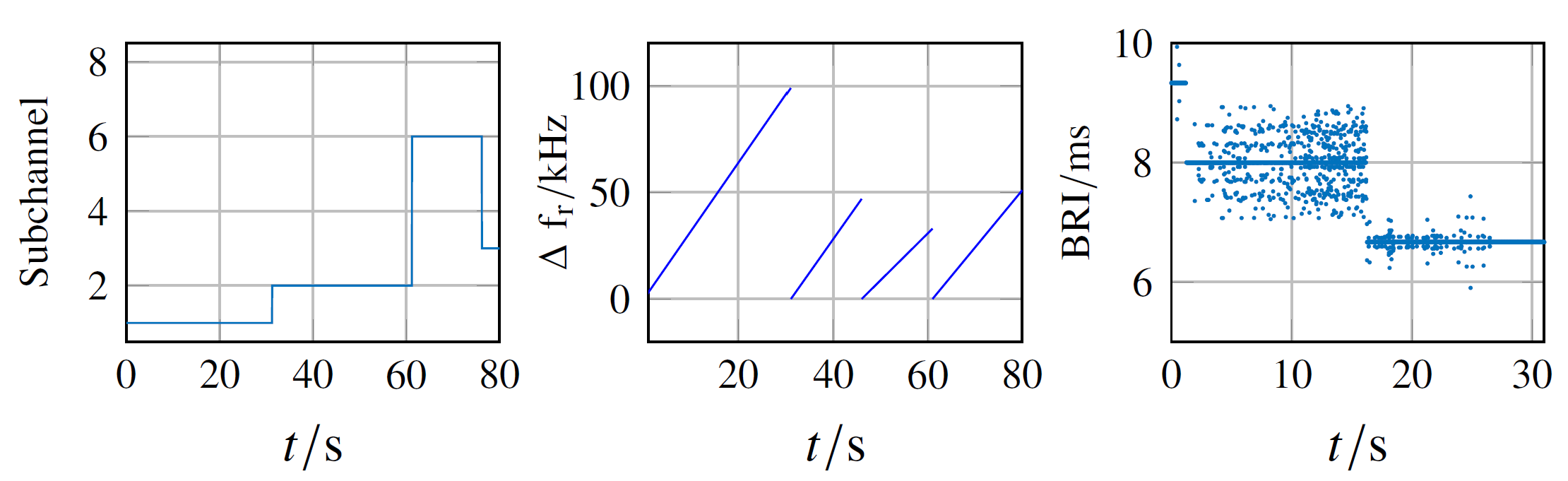}
\caption{Subchannel, relative carrier frequency, and BRI over the time of detection of each burst}
\label{figWholeMeasurement}
\end{figure}

The signal analysis shows, that the vast majority of the bursts (8519 of 8776 revealed bursts) have a bandwidth of around $B_{us} = \SI{62.5}{\mega\hertz}$, which corresponds to one of eight subchannels within the uplink bandwidth $B_u$. It is noticeable that large blocks of consecutive bursts use the same subchannel. The rare but regular changes in the used subchannel could be explained by a handover between different satellites that use different subchannels.
For the Burst Repetition Time (BRI), defined as the time between the beginning of consecutive bursts, of a large majority of the bursts, it seems to apply $\text{BRI} \in \{\SI{6.67}{\milli\second}, \SI{8.00}{\milli\second}, \SI{9.33}{\milli\second}, \SI{10.67}{\milli\second}, \SI{16.00}{\milli\second}, \SI{18.67}{\milli\second} \}$. The subchannels of the bursts with bandwidth $B_{us}$ are shown in figure \ref{figWholeMeasurement}, as well as the BRI of those bursts in subchannel 1 ($\SI{14.0}{\giga\hertz} - \SI{14.0625}{\giga\hertz}$). (Very few bursts with higher BRIs are not shown.) The burst duration seems to be highly variable. While some durations, like, e.g., \SI{0.84}{\milli\second}, are frequently used, the duration seems to be adaptable in timesteps of \SI{17.87}{\micro\second}. The rough estimation of the received carrier frequency $\Delta f_r$ (relative to the frequency of the first burst in the same subchannel) shows that the user terminal applies Doppler shift pre-compensation to the uplink signal.

\paragraph*{Correlation based analysis}
To further investigate the signal, the following correlation algorithm is established: for two complex signals $y_1[n]$ with $n \in \{0, ..., L_{y_1} \}$ and $y_2[n]$ with $n \in \{0, ..., L_{y_2} \}$ and $L_{y_1} > L_{y_2}$ the correlation $r_{y_1,y_2}$ can be calculated with
\begin{equation} \label{corrEquation}
r_{y_1,y_2}[l] = \frac{1}{A} r'_{y_1,y_2}[l] = \frac{1}{A} \sum_{n=0}^{L_{y_1}} y_1[n] y_2^\ast\Big[n-l\Big]
\end{equation}
where $(\cdot)^\ast$ represents the complex conjugate of a complex value. The normalization factor $A$ is given by
\begin{equation} \label{equNormalization}
A = \sqrt{r'_{\hat{y}_{1},\hat{y}_{1}}[0]\ r'_{y_2,y_2}[0]}
\end{equation}
where $\hat{y}_1$ is the part of $y_1$ with length $L_{Y_2}$ that starts with index $\hat{l} = \underset{l}{\text{max}} \{r'_{y_1,y_2}\}$.

The structure of the uplink signal was determined by explorational calculations of (\ref{corrEquation}) with different parts of $s_i$.

\begin{figure}[!b]
\includegraphics[scale=0.45, clip]{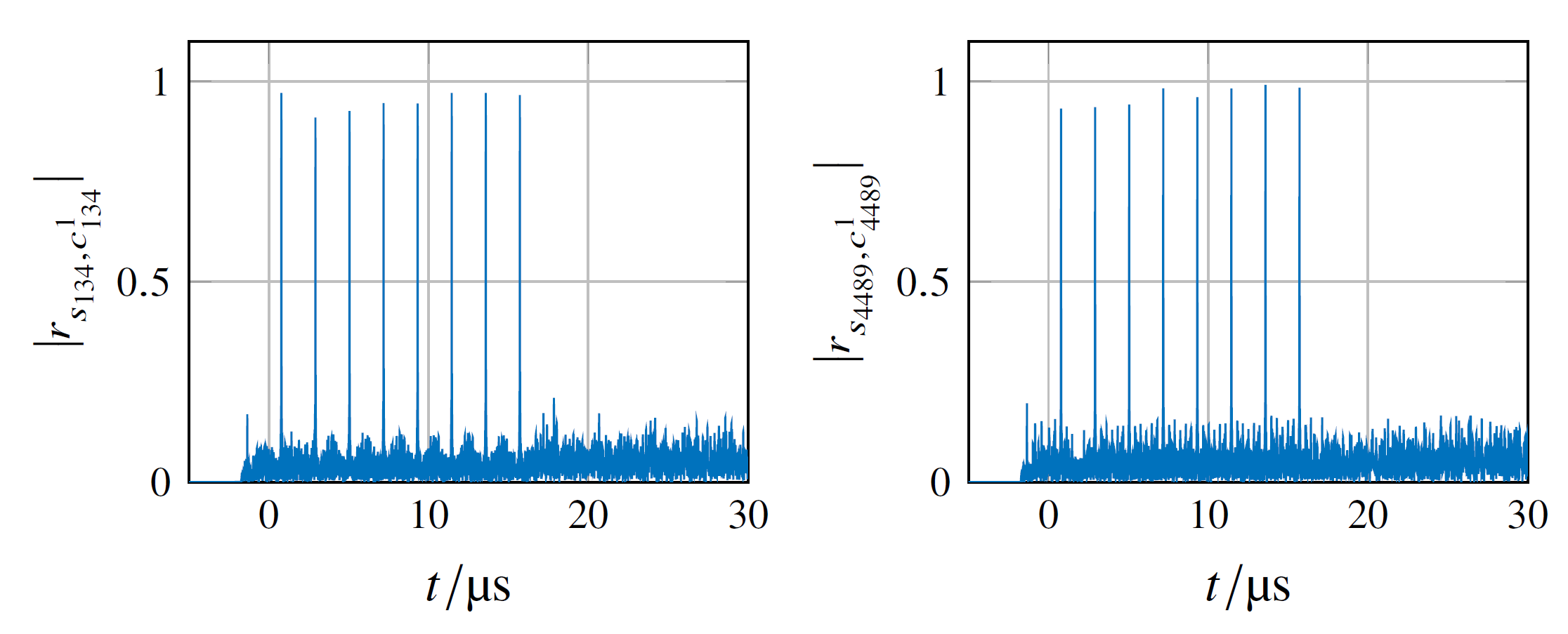}
\setlength{\abovecaptionskip}{-10pt}
\caption{Correlation $r_{s_{i},c_i^1}$ for two exemplary bursts}
\label{figAcfBursts}
\end{figure}

With the given sampling rate $f_s$, the parameters of the structure can be specified as $\dot{L}_c = 1200$ and $\gamma \dot{L}_c = 220$. The correlations in figure \ref{figAcfBursts} between a burst signal $s_i$ and the first element of the synchronisation sequence of the same burst $c_i^1$ were calculated with those values and validate the findings.

Further analysis shows that a small number of different correlation sequences $c$ seem to be used in the uplink. Which specific sequence is used appears to be mainly connected to the burst's BRI and the subchannel and therefore presumably the satellite.
However, more than $70\%$ of the bursts in subchannel 1 seem to use one of three different correlation sequences. 
Additionally, the maximum correlation coefficients between different bursts do not significantly exceed $0.8$ in most cases. This indicates that even for what above is considered to be the same correlation sequence $c$ indeed is not exactly the same sequence. 
Furthermore, in some bursts, the relationship between the phase of the transmitted elements $c^k$ is not as described in the section above and seems unpredictable.

\subsection{Downlink signal properties}
This section sums up the analysis from \cite{Humphreys2022} as far as they are relevant for this work. For Starlink user downlink signals, \SI{2}{\giga\hertz} of bandwidth are allocated. Each Starlink beam uses one of 8 subchannels with bandwidth $B_d = \SI{240}{\mega\hertz}$.
In time domain, the downlink signal is composed of consecutive bursts $s_i$, which are thus called frames, with length $T_f = 1.3\overline{3}\ \si{\milli\second}$. Every frame starts with a synchronisation sequency $c$, which has the exact structure described in the signal model above. Each of the eight subsequences $c^k$ are identical for every burst and satellite, except that the first subsequence and the cyclic prefix are inverse. The subsequences are $T_{c^k} = \SI{4.27}{\micro\second}$ in length, use the entire bandwidth of a subchannel $B_d$, and are made up of 127 (known) DPSK-modulated symbols. The relative length of the prefix $c^p$ is $\gamma = 1/32$. 
The data signal $d_i$ includes 302 OFDM-like symbols. Aside from the synchronisation sequence $c$, the first, last, and (in parts) the second to last of the 302 OFDM-like symbols are constant and appear to be used for synchronisation as well.

\section{Burst detection and frequency estimation} \label{secBurstDetecion}

In this section, an algorithm for Starlink burst detection and a two-step algorithm for frequency estimation are presented. Those algorithms are correlation-based and utilize the synchronization sequence $c$. The properties of the presented algorithms are discussed using the Starlink uplink signal.

\subsection{Burst detection algorithm}
The presented burst detection algorithm adds up the magnitude of eight partial correlations between the received signal $s$ and a representative $\epsilon$. The latter is a measured or reproduced copy of $c_i$, consisting of subsequences $\epsilon^k$ and a prefix $\epsilon^p$.

\begin{equation}
d_{s,\epsilon}[l] = \frac{1}{D} \sum_{k=0}^{7} \Big\lvert d^k_{s,\epsilon}[l] \Big\rvert
\end{equation}

\begin{equation} \label{equSecAlgPart}
d^k_{s,\epsilon}[l] = \sum_{n=0}^{L_s} s[n]\ \epsilon^{k} [ n-(\gamma + k) \hat{L}_{c} - l ]^\ast
\end{equation}

The normalization factor $D$ is defined analogous to $A$ in (\ref{equNormalization}).
A burst is detected at sample $l_j$, if $d_{s,\epsilon}[l_j]$ exceeds a certain threshold and is the maximum value within a certain signal duration $T > L_c T_s$.
\nocite{b2}

\subsection{Frequency estimation algorithm}


The carrier frequency estimation of a burst detected at sample $l_j$ is conducted with a two-step algorithm. In a first step, a raw frequency estimation is calculated by determining the maximum value of $r_{s,\epsilon}[l_j]$ when different frequency shifts $\Delta f$ are applied to $\epsilon$ \cite{Mengali1997}.

\begin{equation} \label{equFreq1}
\tilde{f}_j = \underset{\Delta f}{\text{max}} \big\{ r_{s,\epsilon}[l_j] \big\}
\end{equation}

\begin{equation}
\epsilon_{\Delta f}[n] = \epsilon\ e^{j2\pi \Delta f T_s n}
\end{equation}

In a second step, a fine carrier frequency estimation is performed using results from (\ref{equSecAlgPart}).
\begin{equation} \label{equFreq2}
\hat{f}_j = \frac{1}{2 \pi \dot{L}_c T_s}\ \text{arg}\bigg\{ \sum_{k=1}^{7} \Big( d^{k-1}_{s,\epsilon}[l_j] \cdot  d^{k}_{s,\epsilon}[l_j]^\ast \Big) \bigg\} + \frac{g_j}{\dot{L}_c T_s}
\end{equation}


The correction factor $g_j \in \mathbb{Z}$ accounts for the $\frac{1}{L T_s}$ ambiguity of this algorithm. Results from (\ref{equFreq1}) can be used to calculate $g_j$. The estimator (\ref{equFreq2}) is based on an estimator from \cite{Mengali1997}, which meets the Cramér-Rao bound (CRB).

\subsection{Algorithm analysis with Starlink uplink signals}

\begin{figure}[!t]
\includegraphics[scale=0.45]{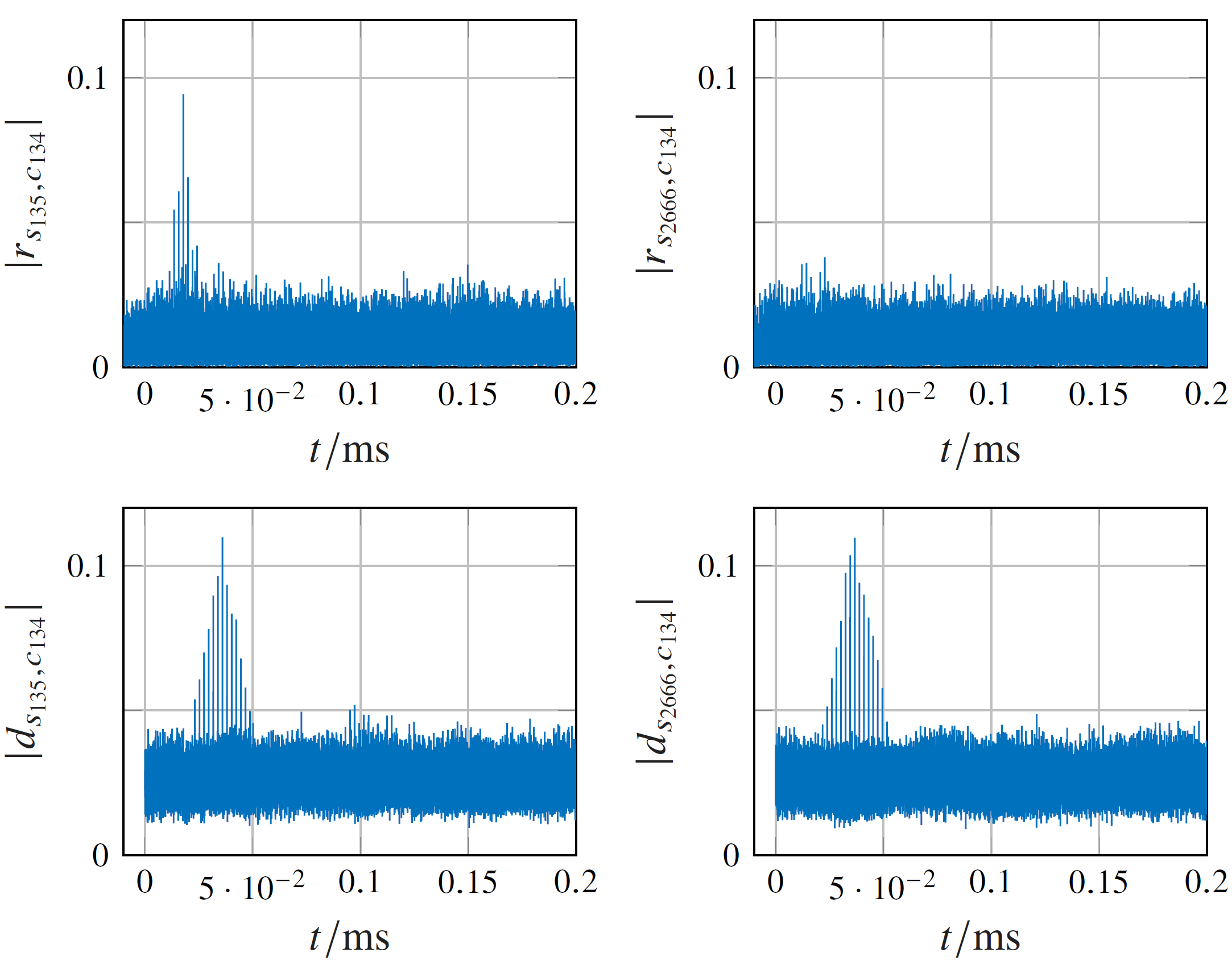}
\caption{Correlation results $r_{s,\epsilon}$ and $d_{s,\epsilon}$ for consecutive  and non-consecutive bursts for $\text{SNR} = \SI{-20}{\decibel}$}
\label{figDetectionResults}
\end{figure}

The presented algorithms are applied to the measured Starlink uplink signal, which contains different synchronization sequences. Therefore, the calculations are conducted with three different representatives $\epsilon_1$, $\epsilon_2$, and $\epsilon_3$.

\begin{figure}[!t]
\includegraphics[scale=0.45]{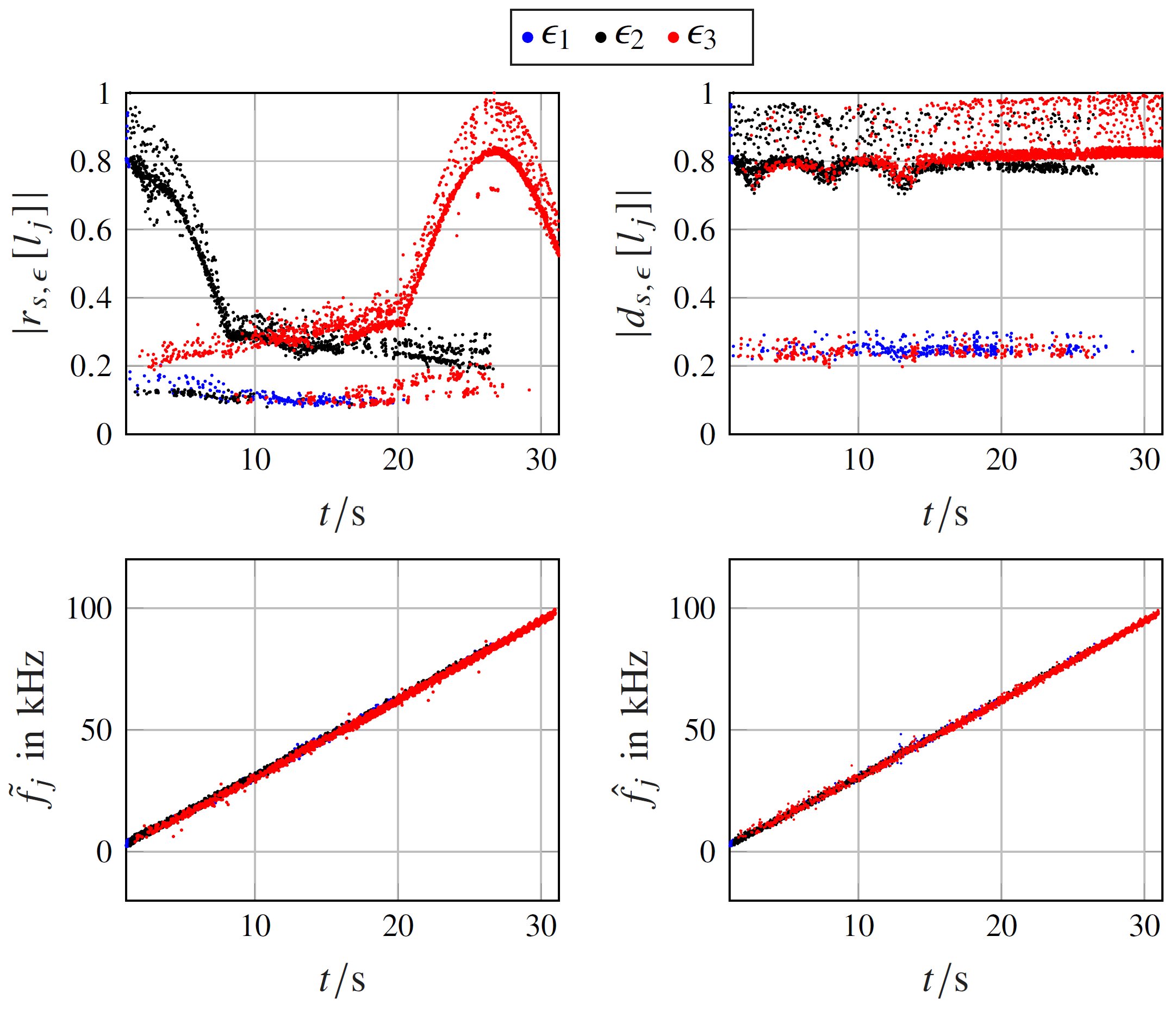}
\caption{Detections with $r_{s,\epsilon}$ and $d_{s,\epsilon}$ and frequency estimations with $\tilde{f}_j$ and $\hat{f}_j$ for all bursts in subchannel 1}
\label{figAllDetAndFreqEst}
\end{figure}

When comparing the presented detection algorithm to a simple correlation-based approach, some properties are noticeable. First, the results $d_{s, \epsilon}$ are significantly less susceptible to an unknown carrier frequency offset or Doppler shift in the received signal than the results of a simple correlation $r_{s, \epsilon}$. This is observed from figure \ref{figDetectionResults}, where the algorithms are calculated for consecutive and non-consecutive bursts. When the burst used as a representative and the burst under investigation are transmitted within a short time period, (nearly) the same Doppler shift pre-compensation is applied to both. Otherwise, significantly different pre-compensations are applied, resulting in a significant decrease in the magnitude of $r_{s, \epsilon}$. The same conclusion can also be derived from figure \ref{figAllDetAndFreqEst}, which shows $r_{s, \epsilon}$ and $d_{s, \epsilon}$ at the samples $l_j$, where bursts are detected. For improved clarity, only the results with the best fitting representative for each burst are presented.
Furthermore, the results in figure \ref{figDetectionResults} show that $r_{s, \epsilon}$ suppresses the (added Gaussian-distributed) noise better due to a higher correlation gain. However, $r_{s, \epsilon}$ is significantly more computationally expensive than $d_{s, \epsilon}$.

Frequency estimation results are presented in figure \ref{figAllDetAndFreqEst}, as well. Again, only the frequency estimations with the representative with the best detection properties ($d_{s, \epsilon}[l_j]$) are considered there. Also, as the representatives are not in baseband, the results contain a frequency offset.

\section{Positioning accuracy estimation} \label{positioningAccuracy}
In the following section, the achievable positioning accuracy is calculated. As a first step, the availabe SNR at the receiver antenna output is estimated. Thereafter, a lower bound of the frequency estimation error is presented. Consequently, the resulting error for Doppler shift based positioning is derived. It is important to mention that additional error sources like, e.g., the ephemeris errors are not considered here.

\paragraph*{Assumptions about the SNR}

With simple transformations of equations from \cite{Molisch2011} the $\text{SNR}_r$ at the receiver antenna output can be calculated with
\begin{equation} \label{equSNR}
\text{SNR}_r = \frac{\Phi_t \lambda_c^2 G_r}{4 \pi k_\text{B} T_N}
\end{equation}
where $\Phi_t$ is the spectral flux density of the transmitted signal, describing the power per surface area and per wavelength, $\lambda_c$ is the carrier wavelength of the downlink signal, and $G_r$ is the receiver antenna gain. The Boltzmann constant is denoted by $k_\text{B}$, the noise temperature by $T_N$.



\paragraph*{Carrier frequency offset estimation}
When estimating the carrier frequency offset $\nu$ of the received signal, the estimation accuracy is lower bounded by the modified Cramér-Rao bound (MCRB) from \cite{Mengali1997} 

\begin{equation} \label{equMCRB}
\text{MCRB}(\nu) = \frac{3}{T_b^2 2 \pi L_0^3} \frac{1}{\text{SNR}_r}
\end{equation}

with the symbol duration $T_b$, and observation duration $L_0 T_s$.

\paragraph*{Positioning estimation}
As a last step, PNT information is calculated from the conducted frequency measurements. Therefore, $N$ measurements with zero-mean, independent, Gaussian distributed measurement errors with variance $\sigma^2$ are assumed. Additionally, a static receiver with known altitude $x_h$, and unknown longitude $x_l$ and latitude $x_b$ (in geodetic coordinates) is assumed. In accordance with \cite{Guo2014}, for this scenario the CRB can be specified as

\begin{equation} \label{PNT_CRB}
\text{CRB}_{x_l,x_b} = \sigma^2 \text{tr}\big( (H^T H)^{-1} \big)
\end{equation}
where $\text{tr}(\cdot)$ represents the trace of a matrix. The matrix $H$ is defined as

\begin{equation} \label{equH}
H = 
\begin{bmatrix}
\frac{\partial f_1(x_l,x_b)}{\partial x_l} & ... & \frac{\partial f_N(x_l,x_b)}{\partial x_l}\\
\frac{\partial f_1(x_l,x_b)}{\partial x_b} & ... & \frac{\partial f_N(x_l,x_b)}{\partial x_b} \\
\end{bmatrix}
^T
\end{equation}

with $f_1(x_l,x_b), ..., f_N(x_l,x_b)$ being the received carrier frequencies (including Doppler shift) at the $N$  time instances the measurements were conducted. 

\paragraph*{Results for tracking a single Starlink satellite}

Equations (\ref{equSNR}-\ref{equH}) are now used to calculate the lower bound of the positioning error. The $\text{MCRB}(\nu)$ from (\ref{equMCRB}) is used as variance $\sigma^2$ in (\ref{PNT_CRB}). The received frequencies $f_1(x_l,x_b), ..., f_N(x_l,x_b)$ are calculated using the spherical-earth model from \cite{Chen2016}, omitting the earth rotation.

The following scenario is assumed: a receiver tracks a single Starlink satellite for the timespan $t_a$ and estimates the frequencies $f_n \in \{f_1, ..., f_N\}$ of the synchronization sequences $c$ at time instances $t_n = q T_f$ with $q \in \{-\frac{N-1}{2}, ..., \frac{N}{2}\}$. $T_f = \frac{1}{750} \si{\second}$ is the repetition time at which Starlink transmits the synchronization sequence.
Assuming that the full synchronization sequence of the Starlink user downlink signal is used, the symbol duration is $T_b = \SI{4.17e-9}{\second}$ and the number of observed symbols is $L_0 = 8 \cdot 127$ \cite{Humphreys2022}.
The satellite passes the zenith at $t = \SI{0}{\second}$ at an orbit with hight $x_{hs} = \SI{550}{\kilo\meter}$. The transmitted carrier frequency is $f_c = \SI{11.7}{\giga\hertz}$, the spectral flux density is assumed to be $\phi_t = \SI{-122}{\decibel\per\meter^2\per\mega\hertz}$. This is the maximum value at the ground according to Starlink's FCC filing \cite{ffcSpaceX}.

\begin{figure}[!t]
\centering
\setlength{\belowcaptionskip}{0pt}
\includegraphics[trim = {1.00cm 0 0 0cm},scale=0.26]{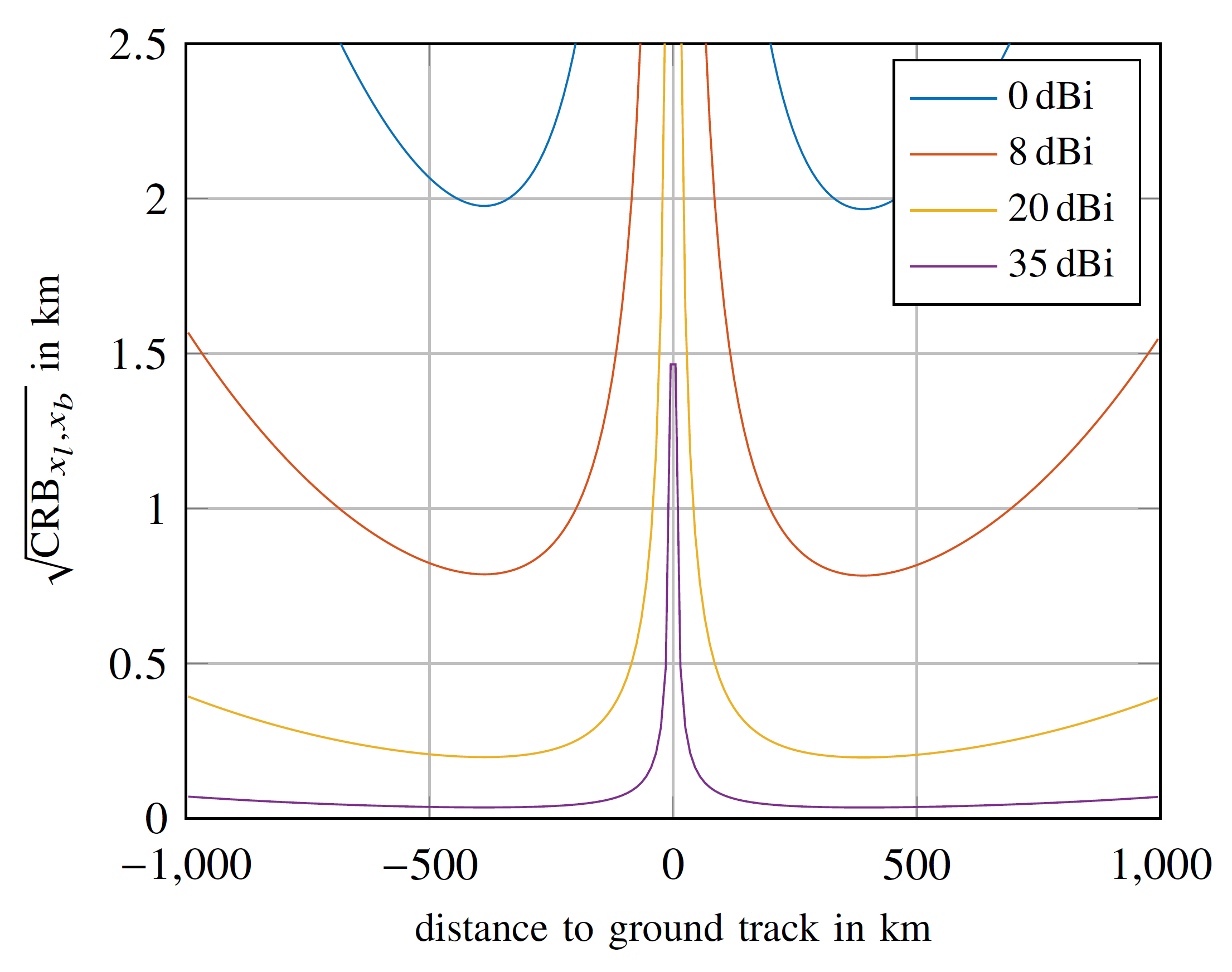}
\caption{Lower bound for the standard deviation of the positioning error for different values of $G_r$ when $t_a = \SI{4}{\minute}$}
\label{figPNTaccuracy}
\end{figure}

\begin{figure}[!t]
\centering
\setlength{\belowcaptionskip}{0pt}    
\includegraphics[trim = {0 0 0.3cm 0}, scale=0.26]{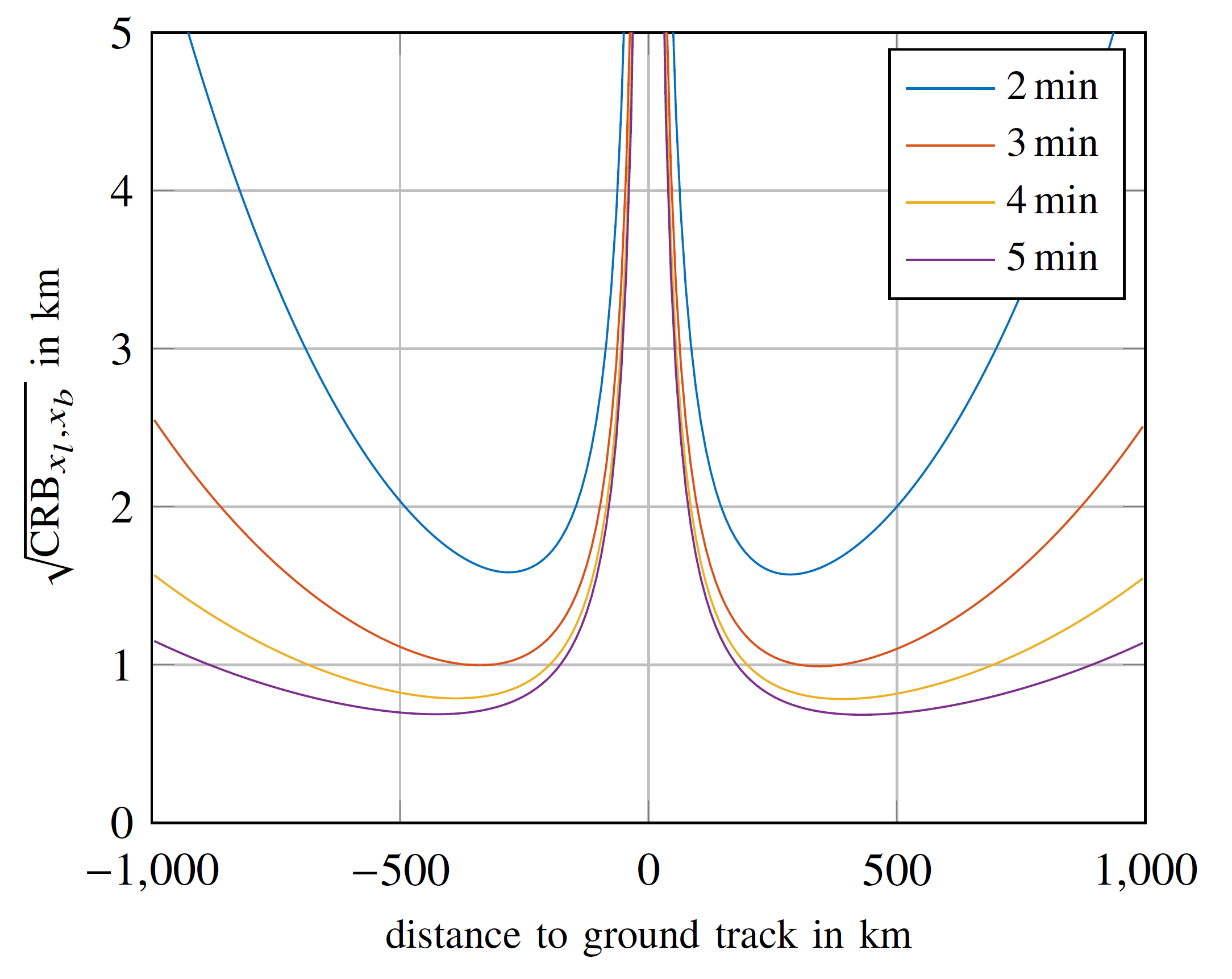}
\caption{Lower bound for the standard deviation of the positioning error for different values of $t_a$ when $G_r = \SI{8}{\decibel}$}
\label{figPNTaccuracy2}
\end{figure}

Figure \ref{figPNTaccuracy} shows the lower bound of the positioning accuracy for the tracking timespan $t_a = \SI{4}{\minute}$ for different receiver antenna gains. For a simple patch antenna with, e.g., $G_r = \SI{8}{\decibel}$, the results show a positioning error of less than \SI{1}{\kilo\meter} for a distance of $200$ - \SI{700}{\kilo\meter} between the receiver and the ground track of the satellite. For smaller distances the accuracy deteriorates rapidely due to inaccuracies in cross-track direction. For larger distances, the estimation in along-track direction is the dominant error source. 
Figure \ref{figPNTaccuracy2} shows the impact of the tracking timespan $t_a$ on the positioning accuracy. Depending on Starlink's beamstearing protocol the maximum timespan to receive and track the main lobe signal of a satellite might be limited.  

\section{Conclusion}
In this work, measurement results of the Starlink user uplink signal are analyzed. Each burst's synchronization sequence is found to consist of 8 repetitions of the same subsequence. Thereby, the identified uplink structure showed significant similarities to the Starlink user downlink signal. Algorithms that utilize the Starlink synchronization sequence for burst detection and frequency estimation are proposed and analyzed by applying them to the Starlink uplink signal.
It is shown that the presented detection algorithm is very robust against an unknown carrier frequency offset or Doppler shift in the received signal. The presented frequency estimation is computationally efficient and promises an estimation variance near the lower bound.
Finally, the impact of frequency estimation errors on the positioning accuracy of Doppler shift based LEO-PNT is investigated by calculating its lower bound. When the Starlink synchronization sequence is used for frequency estimation of a single satellite overflight, the induced positioning errors for most measurement scenarios are in the order of kilometers. Strategies to improve the accuracy include conducting measurements from different satellites with different orbits, using highly directional antennas, and utilizing more parts of the Starlink burst for frequency estimation.



\bibliographystyle{IEEEtran}
\bibliography{IEEEabrv,library}

\end{document}

%% file: figBlockDiagramReceiver.tex
\tikzstyle{int}=[draw, minimum size=2em]
\tikzstyle{init} = [pin edge={to-,thin,black}]

\begin{tikzpicture}[node distance=1.4cm,auto,>=latex']
	\node at (-1.3,0) {\textit{\small \begin{tabular}{l} LEO\\ satellite\\ signals \end{tabular}}};
    \node [int] (b) at (2.4,0) {\begin{tabular}{c} Filtering and \\ downconversion \end{tabular}};
    \node (a) [left of=b,node distance=2.8cm, coordinate] {b};
    	\node at (5.25,-0.35) {\textit{\small \begin{tabular}{l} I/Q\\ data \end{tabular}}};
    \node [int] (e) [below of=a] {\begin{tabular}{c} Burst\\ detection \end{tabular}};
    \node [int] (f) [right of=e, node distance=2.5cm] {\begin{tabular}{c} Frequency\\ estimation \end{tabular}};
    \node [int] (g) [right of=f, node distance=2.6cm] {\begin{tabular}{c} PNT\\ processing \end{tabular}};
    \node [coordinate] (end) [right of=g, node distance=2cm]{};
    \node (z) [below of=g, node distance = 1.2cm] {\textit{\small Ephemerides}};
    \path[->] (a) edge node {$ $} (b);
    \path[->] (e) edge node {$ $} (f);
    \path[->] (f) edge node {$ $} (g);
    \draw[->] (g) edge node {\textit{{\small PNT}}} (end) ;
    \draw[->] (z) edge node {$ $} (g);
	\draw (-0.55,-0.25) arc (-60:60:0.3);
    \draw (3.87,0)--(4.80,0) --(4.8,-0.7) -- (-1.8,-0.7);
    \draw (-1.8,-0.7) -- (-1.8,-1.4);
    \draw[->] (-1.8,-1.4) -- (-1.37,-1.4);
\end{tikzpicture}

%% file: figBlockDiagramMeasurement.tex
\tikzstyle{int}=[draw, minimum size=2em]
\tikzstyle{init} = [pin edge={to-,thin,black}]

\begin{tikzpicture}[node distance=1.4cm,auto,>=latex']
	\node at (-1.3,0) {\textit{\small \begin{tabular}{l} Starlink\\ signal \end{tabular}}};
    \node [int] (b) at (1.1,0) {LNB};
    \node (a) [left of=b,node distance=1.6cm, coordinate] {b};
    \node [int] (c) [right of=b] {FSW};
    \node [int] (d) [right of=c] {IQW};
    \node at (5.4,0) {\textit{\small \begin{tabular}{l} I/Q\\ data \end{tabular}}};
    \node(end) [right of=d, node distance=1.2cm]{};
    \path[->] (a) edge node {\textit{\small \begin{tabular}{c} RF\\ signal \end{tabular}}} (b);
    \path[->] (b) edge node {$ $} (c);
    \path[->] (c) edge node {$ $} (d);

    \path[->] (d) edge node {$ $} (end);
	\draw (-0.65,-0.25) arc (-60:60:0.3);
	
	\draw (0.1,-0.85) circle [radius = 0.25];
	\node (e) at (0.1,-0.87) {$\sim$};
	
    \draw [->] (e) -| (b);
	\draw [->] (e) -| (c);
	\draw [->] (e) -| (d);

\end{tikzpicture}

%% file: main.bbl
\begin{thebibliography}{10}
\providecommand{\url}[1]{#1}
\csname url@samestyle\endcsname
\providecommand{\newblock}{\relax}
\providecommand{\bibinfo}[2]{#2}
\providecommand{\BIBentrySTDinterwordspacing}{\spaceskip=0pt\relax}
\providecommand{\BIBentryALTinterwordstretchfactor}{4}
\providecommand{\BIBentryALTinterwordspacing}{\spaceskip=\fontdimen2\font plus
\BIBentryALTinterwordstretchfactor\fontdimen3\font minus
  \fontdimen4\font\relax}
\providecommand{\BIBforeignlanguage}[2]{{%
\expandafter\ifx\csname l@#1\endcsname\relax
\typeout{** WARNING: IEEEtran.bst: No hyphenation pattern has been}%
\typeout{** loaded for the language `#1'. Using the pattern for}%
\typeout{** the default language instead.}%
\else
\language=\csname l@#1\endcsname
\fi
#2}}
\providecommand{\BIBdecl}{\relax}
\BIBdecl

\bibitem{b4}
Z.~M. Kassas, J.~Khalife, A.~Abdallah, and C.~Lee, ``I am not afraid of the
  jammer: Navigating with signals of opportunity in gps-denied environments,''
  \emph{Proceedings of the 33rd International Technical Meeting of the
  Satellite Division of The Institute of Navigation (ION GNSS+ 2020)}, pp.
  1566--1585, 10 2020.

\bibitem{space2022}
\BIBentryALTinterwordspacing
M.~Wall, ``Watch spacex launch 51 starlink internet satellites on jan. 15 after
  delays.'' [Online]. Available:
  \url{https://www.space.com/spacex-launch-starlink-group-2-4}
\BIBentrySTDinterwordspacing

\bibitem{Neinavaie2021}
M.~Neinavaie, J.~Khalife, and Z.~M. Kassas, ``Exploiting starlink signals for
  navigation: First results,'' in \emph{Proceedings of the 34th International
  Technical Meeting of the Satellite Division of The Institute of Navigation
  (ION GNSS+ 2021)}, 10 2021, pp. 2766--2773.

\bibitem{Neinavaie2022b}
------, ``Acquisition, doppler tracking, and positioning with starlink leo
  satellites: First results,'' \emph{IEEE Transactions on Aerospace and
  Electronic Systems}, vol.~58, pp. 2606--2610, 6 2022.

\bibitem{Neinavaie2022a}
M.~Neinavaie, Z.~Shadram, S.~Kozhaya, and Z.~M. Kassas, ``First results of
  differential doppler positioning with unknown starlink satellite signals,''
  in \emph{2022 IEEE Aerospace Conference (AERO)}.\hskip 1em plus 0.5em minus
  0.4em\relax IEEE, 3 2022, pp. 1--14.

\bibitem{Khalife2022}
J.~Khalife, M.~Neinavaie, and Z.~Z. Kassas, ``The first carrier phase tracking
  and positioning results with starlink leo satellite signals,'' \emph{IEEE
  Transactions on Aerospace and Electronic Systems}, vol.~58, pp. 1487--1491, 4
  2022.

\bibitem{Humphreys2022}
T.~E. Humphreys, P.~A. Iannucci, Z.~Komodromos, and A.~M. Graff, ``Signal
  structure of the starlink ku-band downlink,'' \emph{arXiv}, 10 2022.

\bibitem{NORAD}
\BIBentryALTinterwordspacing
{North American Airospace Defense Command (NORAD)}, ``Two-line element sets.''
  [Online]. Available: \url{http://celestrak.org/NORAD/elements/}
\BIBentrySTDinterwordspacing

\bibitem{Middlestead2017}
R.~W. Middlestead, \emph{Digital Communications with Emphasis on Data
  Modems}.\hskip 1em plus 0.5em minus 0.4em\relax Hoboken, NJ, USA: John Wiley
  \& Sons, Inc., 3 2017.

\bibitem{Psiaki2021}
M.~L. Psiaki, ``Navigation using carrier doppler shift from a leo
  constellation: Transit on steroids,'' \emph{NAVIGATION}, vol.~68, pp.
  621--641, 9 2021.

\bibitem{Kassas2021}
Z.~Kassas, M.~Neinavaie, and J.~Khalife, ``Enter leo on the gnss stage:
  Navigation with starlink satellites,'' \emph{Inside GNSS}, pp. 42--51, 11
  2021.

\bibitem{b2}
C.~A. Hofmann and A.~Knopp, ``Ultranarrowband waveform for iot direct random
  multiple access to geo satellites,'' \emph{IEEE Internet of Things Journal},
  vol.~6, no.~6, pp. 10\,134--10\,149, 2019.

\bibitem{Mengali1997}
U.~Mengali and A.~N. D’Andrea, \emph{Synchronization Techniques for Digital
  Receivers}.\hskip 1em plus 0.5em minus 0.4em\relax Boston, MA: Springer US,
  1997.

\bibitem{Molisch2011}
A.~F. Molisch, \emph{Wireless Communications}, 2nd~ed.\hskip 1em plus 0.5em
  minus 0.4em\relax Wiley Publishing, 2011.

\bibitem{Guo2014}
F.~Guo, Y.~Fan, Y.~Zhou, C.~Xhou, and Q.~Li, \emph{Space Electronic
  Reconnaissance: Localization Theories and Methods}.\hskip 1em plus 0.5em
  minus 0.4em\relax Wiley, 6 2014.

\bibitem{Chen2016}
X.~Chen, M.~Wang, and L.~Zhang, ``Analysis on the performance bound of doppler
  positioning using one leo satellite,'' \emph{2016 IEEE 83rd Vehicular
  Technology Conference (VTC Spring)}, pp. 1--5, 5 2016.

\bibitem{ffcSpaceX}
{FCC filing}, ``Spacex non-geostationary satellite system, attachment a,
  technical information to supplement schedule s,'' 2018.

\end{thebibliography}
